\documentstyle[twocolumn,prl,aps,epsfig]{revtex}
\newcommand{\beq}{\begin{eqnarray}}
\newcommand{\eeq}{\end{eqnarray}}
\renewcommand{\vec}[1]{{\mathbf{#1}}}
\begin{document}
\draft
%\preprint{dvi file made on \today}

\title
{Hall Conductivity near the z=2 Superconductor-Insulator
Transition in 2D}
\author{ Denis Dalidovich$^1$ and Philip Phillips$^{2,3}$}\vspace{.05in}

%
%\begin{instit}
\address
{$^1$National High Field Magnetic Laboratory,\\
Florida State University, Tallahassee, Florida, 32310\\
$^2$Loomis Laboratory of Physics,\\
University of Illinois at Urbana-Champaign,\\
1100 W.Green St., Urbana, IL, 61801-3080\\
$^3$James Franck Institute and Dept. of Physics, University of Chicago\\
5640 S. Ellis Ave., Chicago, Il. 60637\\}

%\end{instit}
%

\address{\mbox{ }}
\address{\parbox{14.5cm}{\rm \mbox{ }\mbox{ }
We analyze here the behavior of the Hall conductivity $\sigma_{xy}$
near a $z=2$ insulator-superconductor quantum critical point in a 
perpendicular magnetic field. We show that 
the form of the conductivity is sensitive to the presence of dissipation
$\eta$, and depends non-monotonically on $H$ once $\eta$ is weak enough.
$\sigma_{xy}$ passes through a maximum at $H \sim \eta T$ in the
quantum critical regime, suggesting that 
the limits $H \rightarrow 0$ and $\eta \rightarrow 0$ do not commute.
}}
\address{\mbox{ }}
\address{\mbox{ }}

%\begin{multicols}{2}
%\twocolumn
%\columnseprule 0pt \narrowtext
\maketitle

Remarkable recent experiments\cite{mk3,rimberg} on the 
insulator-superconductor transition in thin films have explored the 
role of coupling the 2D electron 
gas to a ground plane. Such coupling not only provides a source of 
dissipation but also breaks particle-hole symmetry because the 
uniformly-distributed frustrating offset charges in the ground plane 
cannot be eliminated by Cooper pair tunneling. Mason and 
Kapitulnik\cite{mk3} observed that this upgrade promotes superconductivity 
driving the system closer to phase coherence while at the same time 
the insulating state becomes more insulating. As it is known that 
dissipation\cite{chak} can diminish phase fluctuations, enhancement 
of superconductivity is expected as a result of coupling to a ground plane, 
as is seen experimentally.  However, precisely how dissipation enhances the 
insulating nature of the insulating state is not known.

The issue of the insulating state aside, the inclusion of a ground plane can 
also be used to explore the role of particle-hole symmetry breaking. 
The obvious experimental probe of particle-hole symmetry breaking is the 
Hall conductivity. Only when such symmetry is broken does the Hall 
conductivity acquire a non-zero value. Experimentally\cite{goldman}, 
a non-zero Hall coefficient in thin films exhibiting the IST
has never been observed.  This suggests that in all extant experiments,
particle-hole symmetry is present.  Hence, it would be 
of utmost importance if the experimental set-up with a ground plane 
is used to measure the Hall coefficient. Such measurements would be 
instrumental in delineating how particle-hole asymmetry leads to a 
non-zero Hall coefficient. However, currently no theory exists for the 
Hall coefficient in the vicinity of the IST quantum critical point.  
It is the formulation of the Hall coefficient near the IST that we 
develop here.

The inclusion of particle-hole symmetry breaking results in a fundamental 
change of the dynamical exponent from 
$z=1$ to $z=2$. This can be seen immediately from the following argument.
In a quantum rotor model, the charging term in the presence of offset charges
per rotor is of the form, $E_C(\hat n_i-n_0)^2$, where $E_C$ is 
the charging energy and $\hat n_i$ is the number operator per rotor.
The offset charges enter through the constant term $n_0$.  In the 
corresponding action, the linear term in $n_0$ will be paired with a
linear time derivative with respect to the 
phase, $\partial_\tau\theta$.  As this term will provide the dominant
frequency dependence, the time derivatives will count twice as much as
will their spatial counterparts.  Hence, $z=2$. van Otterlo\cite{otterlo} 
et. al. have outlined an approach to calculate the transverse 
conductivity, $\sigma_{xy}$ based on a Ginzburg-Landau (GL) approach.  
However, these authors did not include dissipation, either internal or 
external. In this brief note, we calculate the fluctuation Hall 
conductivity near the IST in the presence of dissipation, $\eta$. 
The particle-hole asymmetry is assumed to be strong, so that the system 
belongs to the $z=2$ universality class. Transport in the absence of 
magnetic field was examined in Ref. (\cite{z2den}) where 
it was shown that internal dissipation arising from the mutual scattering 
of quasiparticles is exponentially small at low temperatures. Hence, we 
expect that the dominant source of dissipation will arise from the coupling to 
a ground plane. We show that the fluctuation Hall conductivity is large once  
dissipation is weak enough, $\eta \ll 1$. In the quantum critical (QC) regime, 
for $H \ll \eta T$, the Hall conductivity is proportional
to $H$ and depends more singularly on $\eta$ than the longitudinal one.
For larger $H$ the conductivity behaves as $1/H$, but is 
independent of $\eta$ in agreement with the results 
of Ref. (\cite{otterlo}). This leads 
us to the conclusion that for $\eta \ll 1$ the dependence of the 
Hall conductivity on $H$ is non-monotonic.
We emphasize that we consider the fluctuation contribution to the
Hall conductivity on the insulating side of the IST at low temperatures.
In particular non-perturbative effects of a magnetic field are not included.
Hence, we are not concerned here with the
issue of the destruction of the superconducting phase by a finite magnetic
field, in which case the relevant physics is governed by dissipative
motion of field-induced vortices\cite{kapitul}.

The general form of the GL functional that models the behavior near 
the IST point in the presence of particle-hole
asymmetry is\cite{otterlo}
\beq\label{action}
F[\psi]&=&\int d^2r\int d\tau\left\{
\left[\left(\nabla+\frac{ie^*}{\hbar}\vec A(\vec r,\tau)\right)
\psi^*(\vec r,\tau)\right]\right.\nonumber\\
&&\left.\cdot
\left[\left(\nabla-\frac{ie^*}{\hbar}\vec 
A(\vec r,\tau)\right)\psi(\vec r, \tau) \right] + 
\lambda \psi^{\ast}(\vec r,\tau) \partial_\tau\psi(\vec r,\tau)
\right.\nonumber\\
&&\left.+ \kappa^2\left|\partial_\tau\psi(\vec r,\tau)\right|^2
+\delta \left|\psi(\vec r,\tau)\right|^2 + 
\frac{u}{2}\left|\psi(\vec r,\tau)\right|^4 \right\} \nonumber\\
&& +F_{\rm dis}
\eeq
where $\vec{A} (\vec r,\tau)$ is the vector potential due to the applied 
electric field, $e^*=2e$, and  $\delta$ is proportional to the inverse
correlation length. In Fourier space, the dissipation term,
$F_{\rm dis}=\eta \sum_{\vec k, \omega_n} |\omega_n||\psi(\vec k, \omega_n)|^2$
corresponds to the Ohmic model of Caldeira and Leggett\cite{caldeira,ng}.
The parameters $\kappa$ and $\lambda$ measure the strength 
of quantum fluctuations. We will regard $\lambda$ to be on the order of unity.
Consequently, the term proportional to $\kappa$ is irrelevant, 
and all parameters having the dimensionality of energy can be measured 
in units of $\lambda$. The $z=2$ universality class renders the quartic
interaction in Eq. (\ref{action}) marginally irrelevant, making it possible
to perform all calculations in the critical region with logarithmic 
accuracy\cite{z2den,sachsent}. In two dimensions, the static Hall conductivity
obeys the scaling relation\cite{sondhi}
\beq\label{scalecond}
\sigma_{xy}(\delta ,T, H, u)=
\sigma_{xy}(T(l^*),H(l^*), u(l^*)).
\eeq   
In the momentum-shell RG, $l^*$ is the scale at which the 
effective size of the Kadanoff cell becomes on the order of the 
correlation length. The magnetic field scales trivially as 
$H(l)=H e^{zl}$ and we will assume it to be weak enough. 
This means that at the point when the scaling stops,  $\delta (l^*)=1$ and
$H(l^*) \ll 1$. This allows us to neglect the discreteness of the 
Landau energy levels and to obtain the same one-loop RG equations 
as in the absence of a magnetic field,
\beq\label{delta2}
\frac{d\delta (l)}{dl}=2\delta (l)+f^{(2)}(\delta(l),T(l)) u(l)
\eeq
and
\beq\label{u2}
\frac{du(l)}{dl}=-f^{(4)}(\delta(l),T(l)) u(l)^2,
\eeq
where $f^{(2)}$ and $f^{(4)}$ are some complicated functions that
depend strongly on the relation between $\lambda$ and $\eta$.
In the leading order, however, the scaled parameters $H(l^*)$, $T(l^*)$
and $u(l^*)$ are insensitive to the 
form of the particular functional form\cite{z2den}.
Hence, we obtain within logarithmic accuracy that in the QD regime, 
$l^* =\frac{1}{2} \ln(1/\delta )$, giving   
\beq\label{qdth}
T^*=\frac{T}{\delta}, \quad 
H^*=\frac{H}{\delta}.
\eeq 

In the QC regime, one requires a double logarithmic accuracy to obtain
the leading order, so that 
$l^*=\frac{1}{2} \ln (\frac{1}{T} \ln\ln \frac{1}{T})$, while
\beq\label{qdtu}
T^{*}=\ln\ln \frac{1}{T}, \quad 
H^{*}=\frac{H}{T} \ln\ln \frac{1}{T}.
\eeq
\begin{figure}
\begin{center}
\epsfig{file=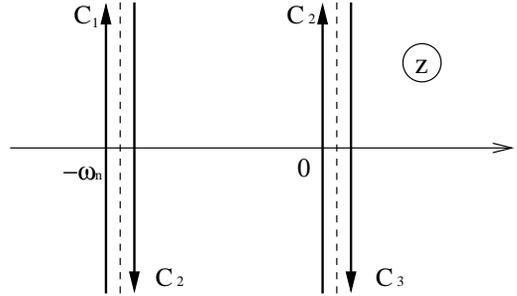, height=4cm}
\caption{The contour of integration in the complex plane $z$. The cuts 
are drawn along the imaginary axis at $z=-\omega_n$ and $z=0$.}
\label{1}
\end{center}
\end{figure}
Sufficiently close to the critical point, the interaction scales to zero.
Hence, we can calculate the conductivity with the help of the Kubo formula,
\beq
\sigma_{xy}(i\omega_n)=-\frac{\hbar}{\omega_n}\int d^2r\int d\tau
\frac{\delta^2\ln Z}{\delta \vec{A}_x (\tau,\bf r)\delta \vec{A}_y (0)}
e^{i\omega_n\tau},\nonumber
\eeq
applied to the Gaussian part of Eq. (\ref{action}) only. A simple calculation 
leads to the result
\beq\label{sigma0}
\sigma_{xy} (i\omega_{\nu}^{*})&=&\frac{i(e^*)^2}
{2h\omega_{\nu}^{*}} T^* (H^{*})^2
\sum_{\omega_n^*} \sum_{n=0}^{\infty} (n+1) \nonumber\\
&&\times \left[ G(\omega_m^* + \omega_{\nu}^* ,n+1) 
G(\omega_m^*, n)\right.\nonumber\\
&&\left. - G(\omega_m^* ,n+1) G(\omega_m^* +\omega_{\nu}^*, n) \right],
\eeq
where 
$G(\omega_m^*, n)=(i\lambda \omega_m^*+\eta |\omega_m^*|+\epsilon_n^*)^{-1}$ 
is the usual 
Matsubara Green function. The rescaled temperature $T^*$ and the energy 
of quasiparticles $\epsilon_n^*=1+H^{*} n$ are employed in the right-hand side
in accordance with Eq. (\ref{scalecond})
($\omega_m^*=2\pi mT^*$). We must perform then an analytical continuation to
real frequencies after doing the summation over $\omega_m^*$.
The latter can be performed by transforming the sum over $\omega_m^*$ 
into the integral over the contour in the complex plane, shown 
on Fig. (\ref{1}). 
The subsequent expansion over the small external frequency 
$\omega =-i\omega_{\nu}$ yields
\beq\label{sigma}
\sigma_{xy} &=&\frac{i(e^*)^2}{8\pi h} (H^{*})^2
\sum_{n=0}^{\infty} (n+1) \int_{-\infty}^{\infty} \coth\frac{z}{2T^*}
\nonumber\\
&&\times \left\{ \left[ G^R_n (z) - G^A_n(z)\right]
\left[ \frac{\partial G^R_{n+1} (z)}{\partial z} +
\frac{\partial G^A_{n+1}(z)}{\partial z} \right] \right.\nonumber\\
&&\left. - \left[ G^R_{n+1}(z) - G^A_{n+1}(z)\right]
\left[ \frac{\partial G^R_n (z)}{\partial z} +
\frac{\partial G^A_n (z)}{\partial z} \right] \right\},
\eeq
where the retarded and advanced Green 
functions
\beq\label{retadv}
G_n^{R/A}=\frac{1}{\lambda z +\epsilon^{*}_{n} \mp i\eta z}
\eeq
have been introduced.

It is evident that the structure of the Hall conductivity is different
from its longitudinal counterpart. In fact, it is not entirely
transparent how the various asymptotic forms can be extracted. 
The situation simplifies, however, for the case of weak magnetic fields
$H^* \ll 1$, the case of interest here. When $H^* \ll \eta$, one can
approximate 
\beq\label{diff}
G_{n+1}^{R/A}(z)-G_{n}^{R/A}(z)=-H^{*} [G_{n}^{R/A}(z)]^2,
\eeq 
and show that the bracketed expression in Eq. (\ref{sigma})
reduces to $-\frac{1}{3}\frac{\partial}{\partial z}[G_n^R (z)-G_n^A (z)]^3$.
Switching then from summation over $n$ to the integration over $t=H^* n$
and integrating subsequently by parts over $z$, we obtain ($z=2T^{*} x$)
\beq\label{sighsmall}
\sigma_{xy} &=&-\frac{8(e^*)^2}{3\pi h} \eta^{3} (T^*)^{3} H^{*}
\int_0^{\infty} t dt \int_{-\infty}^{\infty}\frac{x^3 dx}
{\sinh^2 x}\nonumber\\
&& \times \frac{1}{ \left[ (1+t+2\lambda T^* x)^2 
+4(\eta T^{*}x)^2 \right]^3}.
\eeq
From this expression it is immediately obvious that when
$\lambda=0$ so that particle-hole symmetry is reinstated, the integrand
is an odd function of $x$ and hence vanishes identically when integrated
over the even limits.  This result is expected because particle-hole
asymmetry is essential for the Hall conductivity to be non-zero.

{\it Quantum disordered regime.} In this regime $T^* \ll 1$. 
For weak dissipation, $\eta \ll 1$\cite{z2den,denis}, the
main contributions to the integral over $x$ come from the 
region near $x=0$ and from the vicinity of $x_0 =-(1+t)/2\lambda T^*$.
To obtain the first contribution, $\sigma_{xy}^{(1)}$, we expand the integrand
in Eq. (\ref{sighsmall}) for small $x$. Performing then simple integrations 
and using Eq. (\ref{qdth}), we find that
\beq\label{sig1}
\sigma_{xy}^{(1)}=\frac{128 \pi^3}{225}\frac{e^2}{h}
\frac{\lambda \eta^3 H T^4}{\delta^5}.
\eeq
To calculate the second contribution, we expand the denominator
of the integrand around $x_0$ to arrive at the result,
\beq\label{sig2}
\sigma_{xy}^{(2)}=\frac{e^2}{h} \frac{\lambda^3}{\eta^2}
\frac{HT}{\delta^2} e^{-\delta /\lambda T}, \quad H\ll \eta \delta.
\eeq
The total conductivity for $\eta \ll 1$ in the QD regime 
can be approximately represented
as $\sigma_{xy}=\sigma_{xy}^{(1)}+\sigma_{xy}^{(2)}$. The second contribution
dominates only for very weak dissipation, while for $\eta \sim O(1)$, the
Hall conductivity is given solely by Eq. (\ref{sig1}).

The above derivation is correct assuming the 
condition $H^*\ll \eta$ holds for all $x$. Obviously, near $x_0$ the expansion,
Eq. (\ref{diff}), is not valid if $H\gg \eta$, affecting thus the calculation
of $\sigma^{(2)}_{xy}$. In this case, we introduce 
$y=z+ \epsilon^*_n /\lambda +H^* /2\lambda$ and calculate directly the
difference of the Green functions using Eq.(\ref{retadv}). Expanding then
the cotangent in Eq. (\ref{sigma}) for small $y$ and $H^*$ we obtain with
sufficient accuracy, ($\eta \ll \lambda$)
\beq\label{newsig2}
\sigma_{xy}^{(2)}&=&\displaystyle\frac{4e^2}{\pi h} \frac{\eta^3}{\lambda^4}
\frac{(H^*)^3}{T^*} \sum_{n=0}^{\infty} (n+1) 
\frac{(\epsilon^*)^3}{\sinh^2 (\epsilon^*_n / 2\lambda T^*)}\nonumber\\
&& \times\int_{-\infty}^{\infty} 
\frac{y^2 dy}{\left[ \left( y- \displaystyle\frac{H^*}{2} \right)^2 
+\left( \displaystyle\frac{\eta \epsilon^*_n}{\lambda} \right)^2 
\right]^2}\nonumber\\  
&& \times \frac{1}{\left[ \left( y+ \displaystyle\frac{H^*}{2} \right)^2 
+\left( \displaystyle\frac{\eta \epsilon^*_n}{\lambda} \right)^2 \right]^2}.
\eeq
For $H^* \ll \eta$, one can neglect $H/2$ in the denominator of the equation 
above, and the resultant conductivity reduces to Eq. (\ref{sig2}). In the 
opposite limit $H^* \gg \eta$, the contributions around the minima at
$y=H^* /2$ and $y=-H^* /2$ should be calculated separately yielding
\beq
\sigma_{xy}^{(2)}=\frac{4e^2}{h} \frac{\lambda T}{H}
e^{-\delta /\lambda T}, \quad H\gg \eta \delta.
\eeq
This contribution is $\eta$-independent and inversely proportional to $H$,
representing thus the $\eta \rightarrow 0$ collisionless limit that
was obtained in Ref. (\cite{otterlo}).

{\it Quantum critical regime}. In this regime $T^* \gg 1$ with 
the double-logarithmic accuracy, and the entire contribution is 
determined by small $z$ ($y$). We perform then analogously the integration 
over $y$ in Eq. (\ref{newsig2}) and, using Eq. (\ref{qdtu}), obtain for 
the two limiting cases,
\beq\label{sigqc}
\sigma_{xy}=\frac{e^2}{h}
\left\{ \begin{array}{ll}
       \displaystyle\frac{\lambda^3}{6\eta^2}\frac{H}{T}\left( 
       \ln\ln\frac{1}{T} \right)^2, & \quad
        H\ll \eta T, \\[4mm]
       \displaystyle\frac{4\lambda}{H} \left( \ln\ln\frac{1}{T} \right)
       \left( \ln\ln\ln\frac{1}{T} \right), & \quad
       H \gg \eta T.
       \end{array}\right.
\eeq
The last result in the above formula is written with triple
logarithmic accuracy. 

We see that for small $H$, the conductivity is inversely
proportional to $\eta^2$, while for larger magnetic fields, it does not 
depend on dissipation at all. This corresponds to the existence of a finite 
collisionless limit for the static Hall conductivity. The results here
are different from those for the longitudinal conductance that 
develops the Drude singularity once dissipation is neglected. In the 
collisionless limit, however, $\sigma_{xy}$ is proportional to $1/H$
for all $H$, which has never been observed experimentally
as $H\rightarrow 0$.  Though we limited ourselves with the Gaussian 
approximation, the results obtained indicate that the limits $\eta=0$,
$H\rightarrow 0$ and $H=0$, $\eta \rightarrow 0$ do not commute in the
general scaling formula for the static conductivity near a quantum 
critical point,
\beq
\sigma_{xy}=(4e^2/h)\Sigma_{xy}(\eta /T,\delta^z /T, \sqrt{H}/\delta). 
\eeq
Our results suggest also that for weak dissipation, the dependence of 
$\sigma_{xy}$ on $H$ is non-monotonic, passing through a maximum at
$H \sim \eta T$ in the QC regime. The non-monotonic dependence may be
observed also in the QD regime with the maximum at $H\sim \eta \delta$.
However, in this regime one requires $\eta$ to be so weak 
that $\sigma_{xy}^{(1)}$ is always small, compared to $\sigma_{xy}^{(2)}$.
Experimentally, the suggested dependences can be best tested in 
systems, in which the dissipation is the smallest 
energy scale. 
Such a situation might be realized in artificially fabricated JJA
coupled to a 
ground plane. However, one should remember that the effects of 
magnetic frustration, neglected here, may affect the behavior of
$\sigma_{xy}$ for higher magnetic fields\cite{webb,van der Zant}.
As a result the dependence of the Hall conductivity on magnetic field can
reveal additional minima and maxima. Their origin, however, is not 
connected with dissipation, but rather a consequence of the flux 
quantization. 
   
The calculations presented here are based on the action that describes
also the fluctuations of the superconducting order parameter near a
disorder tuned metal/d-wave superconductor transition\cite{herbut}.
However, $\eta$ is usually of the order of unity in this case, and 
$\sigma_{xy}$ considered here represents only the anomalous 
(fluctuation) part of the total conductivity. 
For $\eta \approx O(1)$ this
anomalous part becomes dominant only unobservably close to the quantum
critical point, in the region where $\ln\ln\frac{1}{T} \gg 1$.
Otherwise, it is of the same order or smaller than the 
normal part and, hence, not interesting. The slow double logarithmic 
divergence of $\sigma_{xy}$ obtained here, is in general agreement with the 
results obtained from the non-linear sigma model approach 
to the IST in a system of interacting bosons in the presence 
of disorder\cite{cham-nayak}. 
  
This work was funded by the ACS PRF Fund.

\end{document}